\documentclass[aps,
 prx,
amsmath,amssymb,
reprint,
floatfix,
numerical,
]{revtex4-1}
\usepackage[colorlinks=true,linkcolor=blue,bookmarks=true]{hyperref}
\usepackage{graphicx}
\usepackage{dcolumn}
\usepackage{bm}
\usepackage[mathlines]{lineno}
\usepackage{float}
\usepackage[utf8]{inputenc}
\usepackage[T1]{fontenc}
\usepackage{mathptmx}
\usepackage[euler]{textgreek}

\begin{document}


\title{Mechanisms of electron-phonon coupling unraveled in momentum and time: The case of soft-phonons in TiSe$_2$}



\author{Martin R. Otto}
\email[]{martin.otto@mcgill.ca}
\affiliation{Department of Physics, Center for the Physics of Materials, McGill University, 3600 rue Universit\'e, Montr\'eal, Qu\'ebec, H3A 2T8}

\author{Jan-Hendrik P\"ohls}
\affiliation{Department of Physics, Center for the Physics of Materials, McGill University, 3600 rue Universit\'e, Montr\'eal, Qu\'ebec, H3A 2T8}

\author{Laurent P. Ren\'e de Cotret}
\affiliation{Department of Physics, Center for the Physics of Materials, McGill University, 3600 rue Universit\'e, Montr\'eal, Qu\'ebec, H3A 2T8}

\author{Mark J. Stern}
\affiliation{Department of Physics, Center for the Physics of Materials, McGill University, 3600 rue Universit\'e, Montr\'eal, Qu\'ebec, H3A 2T8}

\author{Mark Sutton}
\affiliation{Department of Physics, Center for the Physics of Materials, McGill University, 3600 rue Universit\'e, Montr\'eal, Qu\'ebec, H3A 2T8}

\author{Bradley J. Siwick}
\email[]{bradley.siwick@mcgill.ca}
\affiliation{Department of Physics, Center for the Physics of Materials, McGill University, 3600 rue Universit\'e, Montr\'eal, Qu\'ebec, H3A 2T8}
\affiliation{Department of Chemistry, McGill University, 801 rue Sherbrooke Ouest, Montr\'eal, Qu\'ebec, H3A 0B8}



\begin{abstract}
The complex coupling between charge carriers and phonons is responsible for diverse phenomena in condensed matter. We apply ultrafast electron diffuse scattering to unravel electron-phonon coupling phenomena in 1T-TiSe$_2$ in both momentum and time. We are able to distinguish effects due to the real part of the many-body bare electronic susceptibility, $\Re\left[\chi_0(\mathbf{q})\right]$, from those due to the electron-phonon coupling vertex, $g_{\mathbf{q}}$, by following the response of semi-metallic (normal phase) 1T-TiSe$_2$ to the selective photo-doping of carriers into the electron pocket at the Fermi level. Quasi-impulsive and wavevector-specific renormalization of soft zone-boundary phonon frequencies (stiffening) is observed, followed by wavevector-independent electron-phonon equilibration. These results unravel the underlying mechanisms driving the phonon softening that is associated with the charge density wave transition at lower temperatures. 
\end{abstract}


\date{\today}


\maketitle


\section{introduction}

Exotic properties and ordering transitions in quantum materials often arise due to interacting electronic and lattice degrees of freedom that compete for a non-trivial ground state. For example, the onset or suppression of superconductivity can be closely related to the existence of a charge-density wave (CDW) phase~\cite{Chang2012,Zhu2015}. Both phases can emerge from microscopic electron-phonon coupling processes but with vastly different macroscopic properties. To date, the absence of experimental approaches capable of directly probing the relative strength of wavevector (or momentum)-dependent carrier-lattice interactions~\cite{Devereaux2016} and the interplay between the electronic susceptibility ($\chi(\mathbf{q})$) and phonon excitations~\cite{Johannes2008,Chan1973} has profoundly hindered progress in understanding quantum materials. Here we show that ultrafast electron diffuse scattering~\cite{Chase2016,Waldecker2017,Stern2018,Konstantinova2018,deCotret2019,Krishnamoorthy2019,Maldonado2020} (UEDS) provides a direct window on these interactions by unraveling the fundamental mechanisms involved in the zone-boundary transverse phonon softening that is associated with the CDW transition in TiSe$_2$~\cite{Holt2001}. In both time and momentum, UEDS separates the effects of phonon-frequency renormalization resulting from impulsive photocarrier doping (and its associated effect on $\chi(\mathbf{q})$) from the subsequent effects of lattice heating that result from the re-equilibration of electron and lattice systems.  

\begin{figure}[t!]
        \includegraphics[width=0.9\columnwidth]{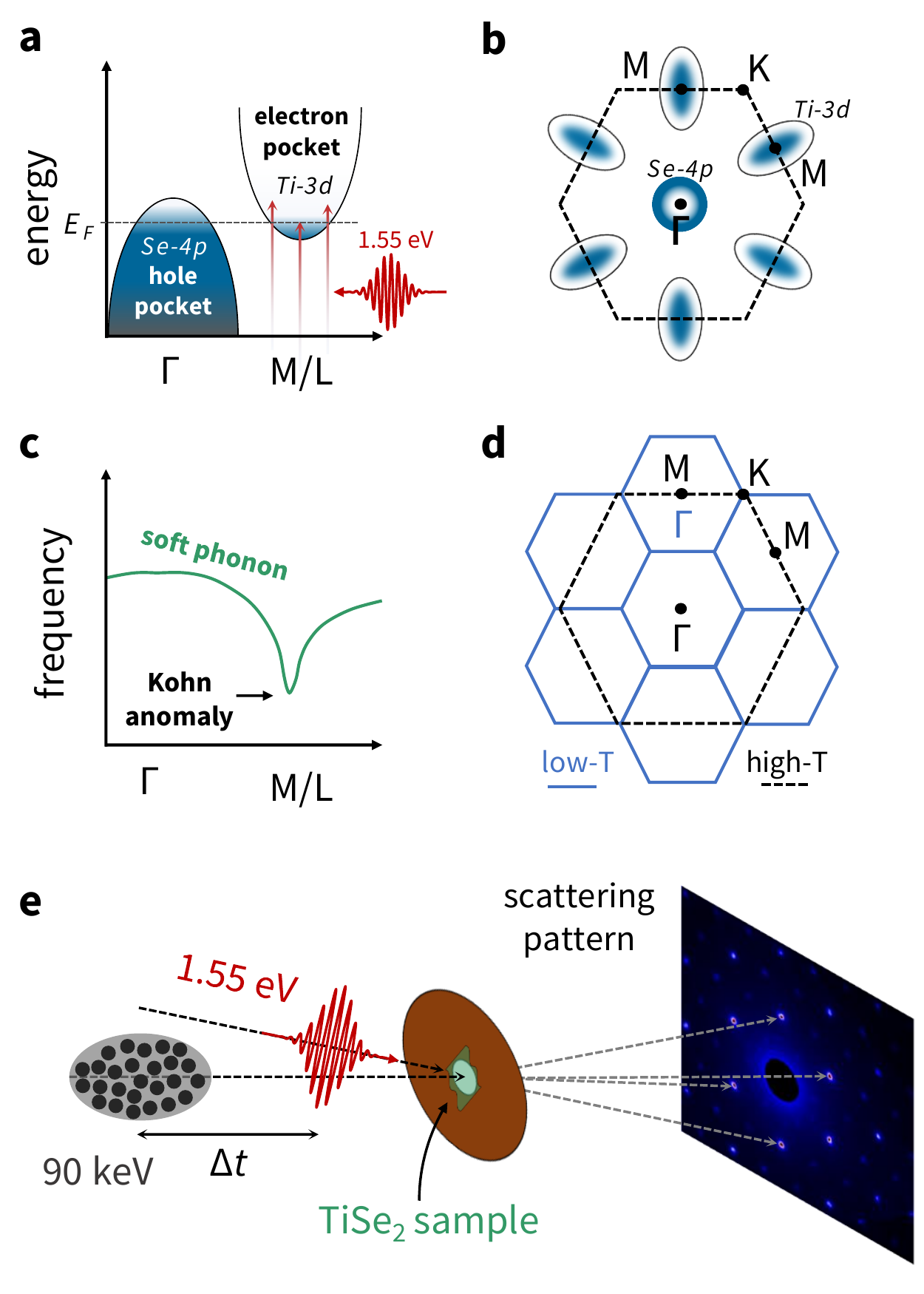}
        \caption{TiSe$_2$ properties and experimental schematic. a) Illustration of valence and conduction bands in TiSe$_2$. 1.55 eV photons drive electronic transitions into a partially occupied conduction band minima at the M-point of the Brillouin zone. The valence (conduction) band at $\Gamma$ (M/L) is formed by Ti--$3d$ (Se--$4p$) orbitals. b) Fermi surface contours showing the bands in a) in the TiSe$_2$ Brillouin zone. c) Dispersion of the transverse phonon in TiSe$_2$ illustrating softening of the frequency at the M and L points yielding a Kohn anomaly. d) Brillouin zone of TiSe$_2$ in the high- ($P\bar{3}m1$) and low- ($P3c1$) temperature phases. Dashed black (solid blue) line illustrates the Brillouin zones of the normal (CDW) phase. The M-points of the high-temperature phase are $\Gamma$-points in the CDW phase. e) Experimental configuration for the ultrafast electron scattering experiments, with the electron beam oriented along the [001] zone axis of TiSe$_2$.}\label{fig:TiSe2_structure}
\end{figure}

A layered transition metal dichalcogenide~\cite{Manzeli2017}, TiSe$_2$ exhibits a rich phenomenology emerging from carrier-lattice interactions~\cite{Calandra2011,Watson2019,Mathias2016}. An indirect semi-metal at room temperature~\cite{Zunger1978,Stoffel1985}~(FIG.~\ref{fig:TiSe2_structure}~a-b)), electron-hole (exciton) pairing~\citep{Jerome1967,vanWezel2010,Monney2015,Pasquier2018} is present in addition to Cooper-pairing~\cite{Morosan2006,Joe2014} and CDW order~\cite{DiSalvo1976}. A commensurate CDW phase forms below $T_c\approx 190~$K that exhibits a $2\times 2\times 2$ superlattice reconstruction~\citep{DiSalvo1976,Brown1980} depicted in FIG.~\ref{fig:TiSe2_structure}~d) This transition is preceded by the observable softening of the entire M--L transverse phonon branch (see FIG.~\ref{fig:TiSe2_structure}~c)) over a temperature range greater than 150 K above $T_c$~\cite{Holt2001}, suggesting that the electron-phonon coupling could play an important role in the emergence of CDW order and the selection of an ordering vector~\cite{Weber2011}. This softening has been investigated by both diffuse~\cite{Holt2001} and inelastic~\cite{Weber2011} X-ray scattering, however, these equilibrium measurements provide limited information on the nature of the microscopic couplings responsible for the observed phonon softening. Recently, static momentum-resolved electron energy loss experiments~\cite{Kogar2017} have also measured the dispersion and softening of a plasmon mode in TiSe$_2$ over a similar temperature range. This work along with other  studies~\cite{Cercellier2007,Monney2011,Monney2012,Monney2016,Hildebrand2016,Mathias2016,Lian2019} point to the strong influence of electron-hole correlations that in turn may drive the CDW transition. This scenario is best understood as an \textit{exciton condensate} predicted over 50 years ago~\cite{Jerome1967}. In the Cu-intercalated species, Cu$_x$Ti$_{1-x}$Se$_2$, CDW order is quenched, yielding a superconductor~\cite{Morosan2006,Calandra2011} which suggests a delicate relationship between the carrier concentration and the lattice stability.

In this article, we present ultrafast electron diffuse scattering (UEDS) measurements on 1T-TiSe$_2$ in the normal phase at 300 K. Our focus is on the fundamental mechanisms that underlie the observed softening of the zone-boundary transverse phonon branch along M-L of the Brillouin zone (BZ).  As mentioned above, this branch softening is associated with the three dimensional CDW transition at lower temperatures, whose ordering vector runs through the L points of the BZ of the normal phase. Figure~\ref{fig:TiSe2_structure}~e) shows a schematic of the experimental geometry. We have previously shown that UEDS provides a momentum-resolved view of phonon dynamics~\cite{Stern2018, deCotret2019} in a pump-probe configuration with $\sim$100~fs time resolution~\cite{Otto2017}.  Here we show that UEDS also allows for the separation of impulsive changes to the real part of $\chi(\mathbf{q})$ induced directly by photo-doping from the subsequent coupling of electronic excitation energy into the phonon system. The former is observed as a strong, wavevector-specific renormalization (stiffening) of the transverse soft-mode at the M and L points, and the latter as a nearly isotropic heating of phonon modes throughout the BZ. We identify no specific strongly coupled phonon modes in TiSe$_2$~\cite{Karam2018} from the perspective of electron-phonon energy transfer (or lattice heating), indicating that a highly anisotropic $\chi(\mathbf{q})$ and its dependence on carrier concentration is the primary driver of the phonon softening and lattice instability in TiSe$_2$.
\section{Electron-phonon coupling and the structured susceptibility}
The renormalization of a phonon frequency $\omega(\mathbf{q})$ due to the coupling between electrons and phonons is determined by the structured electronic susceptibility $\chi(\mathbf{q})$ according the following equation~\cite{Johannes2008,Kaneko2018,Chan1973}
\begin{equation}\label{eqn:phonon_renorm}
\omega^2(\mathbf{q}) = \omega_0^2(\mathbf{q}) - \omega_0(\mathbf{q})\chi(\mathbf{q})\big/\hbar,
\end{equation}
where $\omega_0(\mathbf{q})$ is the bare frequency in the absence of coupling. In Eqn.~\eqref{eqn:phonon_renorm}, $\chi(\mathbf{q})$ is given by 
\begin{equation}\label{eqn:struc_chi}
\chi(\mathbf{q}) = -\frac{2}{N}\sum_{\mathbf{k}}\sum_{a,b}\langle g_{\mathbf{q},ab} \rangle^2\frac{f(\epsilon_{\mathbf{k},a}^{0}) - f(\epsilon_{\mathbf{k-q},b}^{0})}{\epsilon_{\mathbf{k},a}^{0}-\epsilon_{\mathbf{k-q},b}^{0}},    
\end{equation}
where $N$ is the particle number and $\mathbf{k}$ is the electronic wavevector. The electron-phonon coupling vertex, $g_{\mathbf{q},ab}$, describes the rate of inelastic single electron scattering between states of energies $\epsilon_{\mathbf{k},a}^0$ and $\epsilon_{\mathbf{k-q},b}^0$ in bands $a$ and $b$ (respectively) through the simultaneous creation or annihilation of a phonon with wave-vector $\mathbf{q}$. The magnitude of this vertex (or the rate of scattering) depends on the strength of the potential energy modulation experienced by the electron due to lattice displacements associated with phonons of wavector $\mathbf{q}$; i.e. phonon coordinates associated with a large energy modulation have an enhanced $g_{\mathbf{q}}$.      

\begin{figure*}[t!]
        \includegraphics[width=0.9\textwidth]{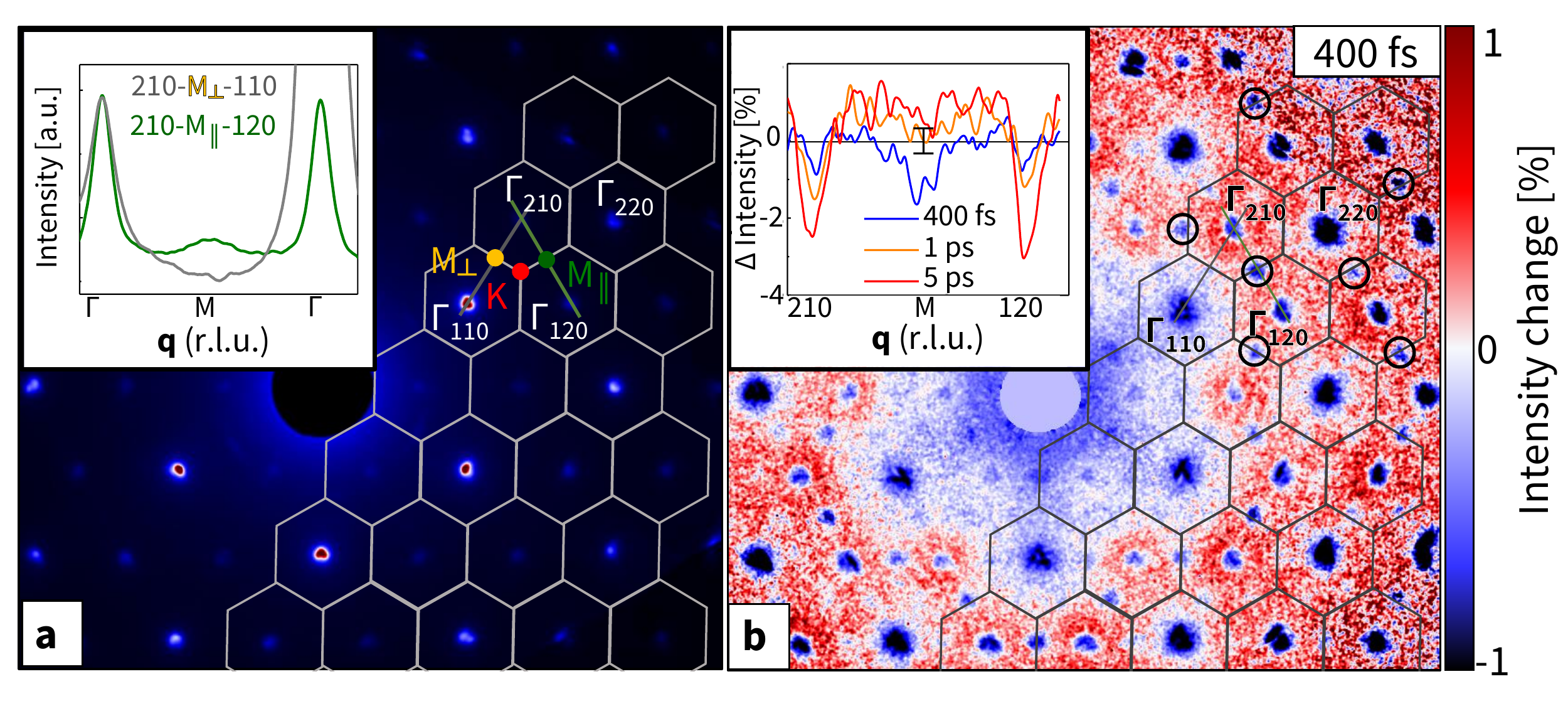}
        \caption{Ultrafast electron scattering of TiSe$_2$. a) Equilibrium electron scattering pattern with various Bragg peaks ($\Gamma$-points) and high-symmetry points (M and K) of the Brillouin zone identified. \emph{Inset:} Intensity line-cuts through M-points along the $\mathbf{a}^*$ (grey) and $\mathbf{b}^*-\mathbf{a}^*$ (green) directions in reciprocal space. The green line-cut intersects a thermal diffuse peak due to a populated transverse acoustic phonon mode. The peak is not present in the grey line-cut in the other direction because of the magnitude of the one-phonon structure factor in Eq.~\eqref{eqn:I1_reduced}. b) Normalized intensity change at a pump-probe time-delay of 400 fs. Regions of decreasing intensity are found not only at the $\Gamma$-points by also at particular M-points where strong TDS intensity from the transverse acoustic mode appears.  These regions are indicated in multiple Brillouin zones by the black circles. These results are in strong agreement with scattering intensity simulations using density function theory results~
        \ref{app:DFT}. \emph{Inset:} Intensity change of green line-cut shown in a) for various time-delays. The noise level of the measurement is indicated by the error bar of 0.2\%.}\label{fig:3_panel_ueds}
\end{figure*}

The occupancy of the electronic states is given by the distribution functions $f(\epsilon_{\mathbf{k},a}^0)$ and $f(\epsilon_{\mathbf{k-q},b}^0)$. The term $f(\epsilon_{\mathbf{k},a}^{0}) - f(\epsilon_{\mathbf{k-q},b}^{0})\big/\epsilon_{\mathbf{k},a}^{0}-\epsilon_{\mathbf{k-q},b}^{0}$ in Eqn.~\eqref{eqn:struc_chi} is the static Lindhard response function and is often called the \textit{bare susceptibility}, $\chi_0(\mathbf{q})$, which differs from the structured susceptibility by the factor $g_{\mathbf{q},ab}$.  The distribution functions $f$ are not restricted to equilibrium Fermi-Dirac statistics, thus the form of Eqn.~\eqref{eqn:struc_chi} is valid for non-equilibrium scenarios. The structured susceptibility $\chi(\mathbf{q})$ is understood to have units of states/eV and reduces $\chi_0(\mathbf{q})$ when $g_{\mathbf{q},ab}$ has an no wavevector dependence. The bare electronic susceptibility describes the linear response of the many-body electron system to lattice potential modulations at wavevector $\mathbf{q}$. It governs the strength of $\mathbf{q}$-dependent \emph{dielectric screening} according to the underlying electronic states at all wavevectors $\mathbf{k}$ along with the availability of states at $\mathbf{k-q}$~\cite{Mahan2013}.

It is worth mentioning that in the ultrafast literature, `electron-phonon coupling' has been used almost exclusively to describe the inelastic scattering processes involved in electron-lattice equilibration following photoexcitation. Such effects are due to $g_{\mathbf{q}}$ (henceforth the $a,b$ band dependence is dropped for simplicity). However, in literature on CDW materials (both theory and experiment) `electron-phonon coupling' more often refers to effects primarily controlled by $\chi(\mathbf{q})$ rather than $g_{\mathbf{q}}$. That is, phenomena related to the dielectric screening of the lattice by carriers, Fermi-surface nesting and the re-normalization of phonon frequencies~\cite{Johannes2008,Zhu2015} (soft modes and structural instabilities). Qualitatively distinct phenomena are described by $\chi_0(\mathbf{q})$ and $g_{\mathbf{q}}$, yet both are often described as electron-phonon coupling. In the remaining sections, we demonstrate how the UEDS technique unravels these qualitatively distinct effects and reveals momentum-dependent electron-phonon coupling in substantial detail.    

\section{experimental results: ueds from photodoped $\textup{TiSe}_2$}

The intensity of first order, thermal-equilibrium diffuse scattering (TDS) at temperature $T$ is given by 
\begin{equation}\label{eqn:TDS_full}
I_1(\mathbf{q}) \propto \sum_j\frac{n_j(\mathbf{q})+\frac{1}{2}}{\omega_j(\mathbf{q})}\left| F_{1j}(\mathbf{q},\hat{\mathbf{e}}_{j})\right|^2,
\end{equation}
where $n_j(\mathbf{q})=\coth\left(\hbar\omega_j(\mathbf{q})/2k_BT\right)$ and $\omega_j(\mathbf{q})$ are the occupancy and frequency of phonon mode $j$ respectively. $F_{1j}(\mathbf{q},\hat{\mathbf{e}}_j)$ is the \textit{one-phonon structure factor} which weights the contribution of phonon $j$ according to the projection of its polarization vector $\hat{\mathbf{e}}_{j}$ onto $\mathbf{q}$ (see Appendix~\ref{app:diffuse_intensity}). For the case of low-frequency phonons ($\hbar\omega\ll k_B T $), Eqn.~\eqref{eqn:TDS_full} simplifies to
\begin{equation}\label{eqn:I1_reduced}
I_1(\mathbf{q})\propto\sum_j\frac{T}{\omega^2_j(\mathbf{q})}|F_{1j}(\mathbf{q},\hat{\mathbf{e}}_{j})|^2.
\end{equation}

For the interpretation of the results that follow it is important to note that in-plane phonon frequencies in TiSe$_2$ are all below $\sim~$9 THz, and that the frequency of phonons along the soft M-L transverse branch are in the 1 - 2 THz (4 - 8 meV) range. Thus, all in-plane modes are thermally populated at 300 K and contribute to the TDS observed before photoexcitation (Eqn.~\eqref{eqn:TDS_full}). The soft-phonons, in particular, are significantly populated. This distinguishes the current experiments from our earlier work on graphite~\cite{Stern2018,deCotret2019}, where the in-plane phonon frequencies are so large, effectively only zero point motion is present in all but the zone-center acoustic modes prior to photoexcitation at 300 K. Here, significant thermal fluctuations of the lattice along all phonon coordinates are present before photoexcitation. TDS intensity provides a measure of the amplitude of these fluctuations at all phonon momenta.  By extension, one expects that UEDS measurements should (in principle) be sensitive to any modulation in the amplitude of these thermal fluctuations that results directly from the photodoping of carriers in addition to the subsequent heating of the lattice through electron-phonon re-equilibration as has been previously shown~\cite{Chase2016,Waldecker2017,Stern2018,Konstantinova2018,deCotret2019}.  

An equilibrium electron scattering pattern of semi-metallic TiSe$_2$ in the normal phase taken along the [001] zone axis is shown in FIG.~\ref{fig:3_panel_ueds}~a).  Hexagons indicate the BZs with Bragg peaks located at the zone centers ($\Gamma$-points).  Also indicated are two different M-points, one between zones 120 and 210 (green) and another between 110 and 210 (orange), along with a K-point (red).  Intensity line-cuts shown in the inset of FIG.~\ref{fig:3_panel_ueds}~a) reveal a TDS peak at $\mathbf{q}=(\frac{3}{2},\frac{3}{2},0)$, the M-point between 120 and 210, produced primarily by a thermally occupied low-frequency transverse phonon. This phonon peak is not observed at $\mathbf{q}=(\frac{3}{2},1,0)$, the M-point between 110 and 210, because $F_{1j}(\mathbf{q},\hat{\mathbf{e}}_{\mathbf{k}j})$ is much smaller (Appendix~\ref{app:DFT}). This definitively demonstrates that this is a phonon TDS peak \emph{not} a weak CDW reflection. We denote M$_{\parallel}$ as the M-points which exhibit a TDS peak in equilibrium and M$_{\perp}$ as those which do not. Qualitatively, the difference between M$_{\parallel}$ and M$_{\perp}$ can be understood by considering that the soft mode is primarily of transverse character; atomic motion is primarily orthogonal to the wavevector at the M-points (\emph{i.e.} orthogonal to the orange/green lines indicated).  Transverse polarization is nearly parallel to $\mathbf{q}$ at the M$_{\parallel}$ point, but nearly orthogonal to $\mathbf{q}$ at the M$_{\perp}$ point which has a strong effect on the dot product in the single phonon structure factor (see Eqn.~\eqref{eqn:TDS_full}) .

The UEDS measurements are carried out in transmission mode at 90 keV in an RF-compressed instrument described in Refs.~\cite{Chatelain2012,Otto2017} and Appendix~\ref{app:experimental_details}.  The sample is photo-excited nearly collinear ($\sim 5^\textup{o}$) with the electron beam illumination. Photo-excitation of TiSe$_2$ at 1.55 eV drives vertical transitions in the M-L region of the BZ~\cite{Zunger1978,Fang1997,Reshak2003,Rohde2014}, effectively photo-doping additional carriers into the electron pockets near the Fermi level (see FIG.~\ref{fig:TiSe2_structure}~b)). Following this photo-excitation, we measure the non-equilibrium dynamics in the phonon system through the normalized intensity changes $\Delta\bar{I}(\mathbf{q},t)=\frac{I(\mathbf{q},t)-I(\mathbf{q},0)}{I(\mathbf{q},0)}$ as a function of pump-probe time delay $t$ where $I(\mathbf{q},0)$ is the equilibrium scattering pattern. Figure~\ref{fig:3_panel_ueds}~b) shows $\Delta \bar{I}(\mathbf{q},t)$ at $t=400$ fs. Immediately evident is the anticipated reduction in Bragg peak (inset, FIG.~\ref{fig:3_panel_ueds}~b)) intensities at the $\Gamma$-points from the Debye-Waller effect.  In addition, however, is the striking and surprising intensity \textit{decrease} found at the M$_{\parallel}$-points where strong TDS signal from the transverse phonon is found (indicated with circles in several highlighted BZ). Line-cuts from 120--M--210 for various time delays (inset, FIG.~\ref{fig:3_panel_ueds}~b)) indicate that the overall negative $\Delta\bar{I}(\mathbf{q}=\mathrm{M}_{\parallel},t)$ lasts only for $\sim1$ ps, yet the relative suppression is even stronger than those of the neighbouring Bragg peaks (FIG.~\ref{fig:3_panel_ueds}~b)). This quasi-impulsive suppression of diffuse intensity at M$_{\parallel}$-points is followed by a rise similar to other points of the BZ. The intensity remains approximately constant (steady-state) for time-delays beyond 5 ps. The dynamics at $\Gamma=120,210$ continue to decrease beyond 1 ps as expected from overall increasing the Debye-Waller factor from lattice heating and phonon anharmonic decay processes~\cite{Stern2018}. 

We investigate these data further by comparing the complete time-dependence at various points in the BZ. Figure~\ref{fig:time_traces} shows UEDS intensity dynamics at the M-, K-, and $\Gamma$-points shown in FIG.~\ref{fig:3_panel_ueds}~a). The $\Gamma=110$-point exhibits a single-exponential dependence involving a $1.09\pm0.03~$ps time-contant. This Bragg peak Debye-Waller behavior is reproduced at all $\Gamma$-points of the scattering pattern and describes the average increase in the mean-square vibrational amplitude of Ti and Se atoms due to the differential phonon excitation across all branches. This behavior is very similar to the diffuse intensity dynamics measured at M$_{\perp}$ and K (FIG.~\ref{fig:time_traces}) which report the transient phonon occupancies at that $\mathbf{q}$ the BZ as expected from Eqn.~\eqref{eqn:I1_reduced}.  The intensity dynamics found at M$_{\parallel}$ are fit to a bi-exponential model convolved with a Gaussian instrument response function (IRF) with full-width-at-half maximum of 130 fs.  The fitting results indicate an initial drop occurring with a time-constant of $109\pm21$ fs followed by a $643\pm110$ fs rise in scattering intensity. The K and M$_{\perp}$ intensities are fit to single exponentials and have $1267\pm189$~fs and $976\pm295$~fs time constants respectively.

\section{Phonon Renormalization}\label{sec:phonon_renorm}

\begin{figure}[t!]
        \includegraphics[width=0.85\columnwidth]{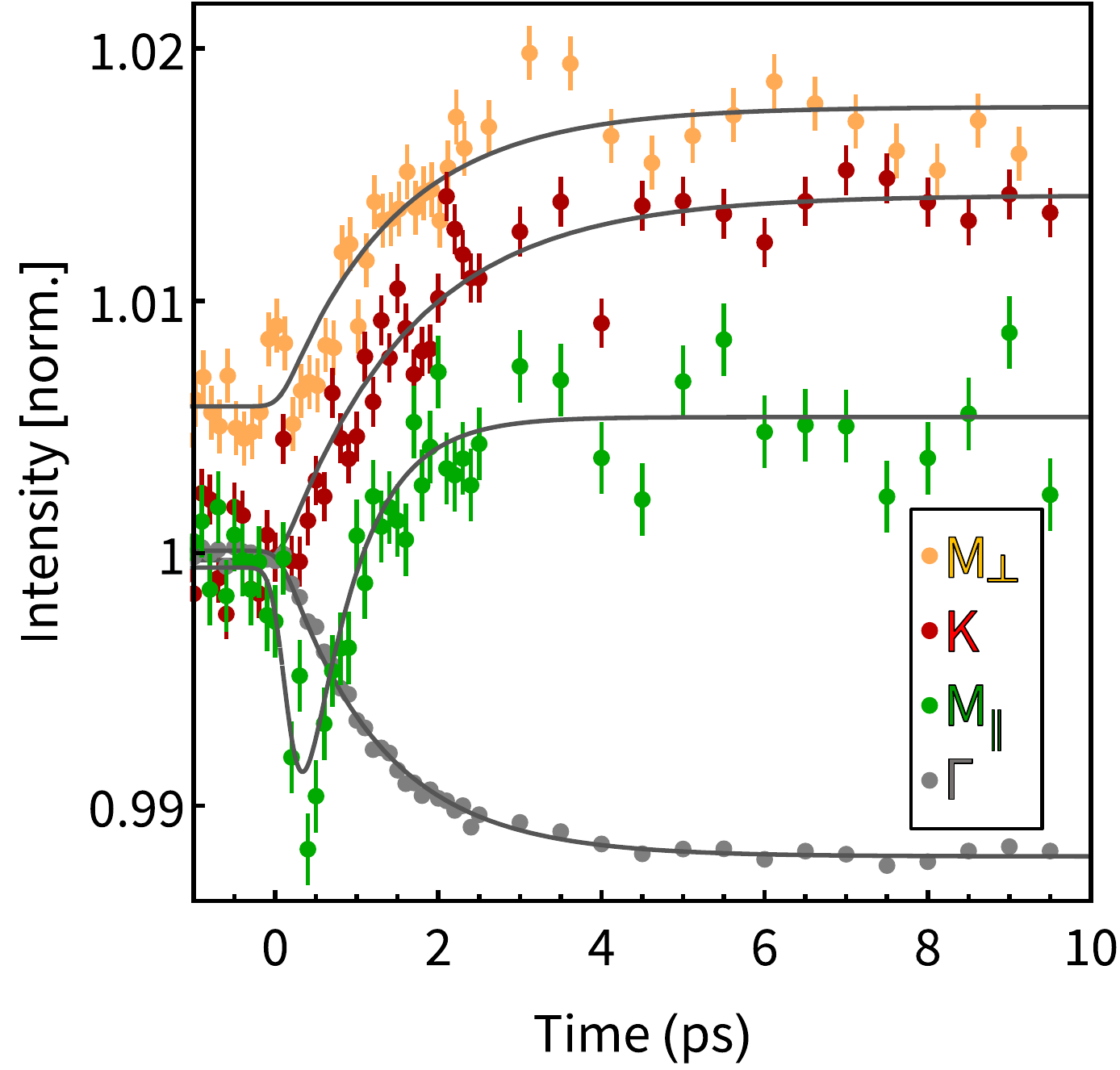}
        \caption{Ultrafast electron diffuse intensity dynamics at various points of the Brillouin zone along with the $\Gamma_{110}$ (Bragg peak) Debye-Waller dynamics. The M$_{\perp}$ trace is shifted for clarity. The $\Gamma_{110}$ trace is scaled by a factor of 1/4. The error bars are determined from the statistics of the intensity before photo-excitation ($t=0$).}\label{fig:time_traces}
\end{figure}

Inspection of Eqn.~\eqref{eqn:TDS_full} indicates that a decrease in $I_1(\mathbf{q}=\mathrm{M}_{\parallel})$ could result from either a reduction in $n_j$ (effective cooling of phonons $j$) or an increase in frequency $\omega_j(\mathbf{q=\textup{M}})$. Photo-excitation at 1.55 eV drives direct/vertical transitions into the electron pocket at the M and L points~\cite{Zunger1978,Rohde2014}, not indirect transitions, and deposits significant electronic energy (0.1 eV per unit cell) into the material.  Given these facts, the unlikely scenario of impulsive cooling of specific phonons can entirely be ruled out. Therefore, we attribute the quasi-impulsive anisotropic suppression in diffuse intensity at M$_{\parallel}$ (FIG.~\ref{fig:3_panel_ueds}) and FIG.~\ref{fig:time_traces}) and L (Appendix~\ref{app:GMGL}) to the re-normalization (stiffening) of the zone-boundary (M/L) transverse phonon mode frequency $\omega_{\textup{T}}(\mathbf{q}=\textup{M})$. This phenomenon can be distinguished from the  heating of phonon modes throughout the BZ (FIG.~\ref{fig:time_traces}), which is observed as an increase in diffuse intensity an order of magnitude slower. 

\begin{figure*}[t!]
        \includegraphics[width=0.95\textwidth]{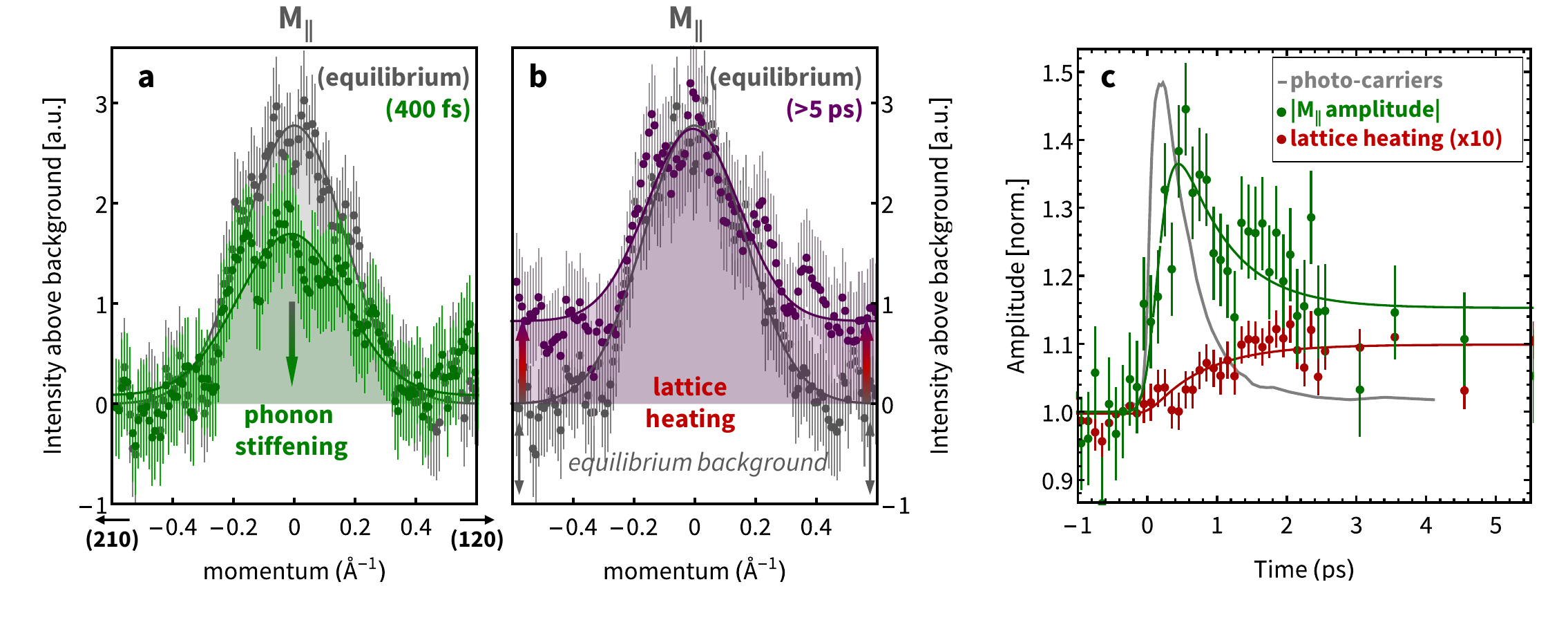}
        \caption{Time-resolved soft-mode scattering in TiSe$_2$. a) Diffuse scattering from the transverse soft-mode at $\mathbf{q}=(\frac{3}{2},\frac{3}{2},0)$ (M$_{\parallel}$) shown under equilibrium conditions (gray) and at a pump-probe delay of 400 fs (green) illustrating a suppression of the peak amplitude due to phonon stiffening. b) Scattering intensity after 5 ps, also with the equilibrium data from a), where the dominant effect is the increased diffuse background due to lattice heating. The peaks are fit at all pump-probe time-delays to extract the time-dependent amplitudes and diffuse background offsets. c) Fit results for M$_{\parallel}$ amplitude (absolute value shown) and diffuse background versus time. The background amplitude is scaled by a factor of 10 for presentation (the actual background rise at late times in roughly 1\% consistent with FIG.~\ref{fig:3_panel_ueds} and \ref{fig:time_traces}). Photo-carrier density at M determined from time and angle resolved photo-electron spectroscopy~\cite{Monney2016} is shown as the gray curve. The error bars in both a) and b) are determined by intensity counting statistics and the standard error in c) is determined from the fitting routine covariance matrix.}\label{fig:Mpeak_results}
\end{figure*}

Equation~\eqref{eqn:phonon_renorm} and \eqref{eqn:struc_chi} describe the phonon renormalization behaviour we observe. The UEDS data for M$_{\parallel}$ shown in FIG.~\ref{fig:time_traces} provides a clear demonstration of both $\chi(\mathbf{q})$ and $g_{\mathbf{q}}$ related electron-phonon coupling phenomena. The initial $109\pm21~$fs intensity decrease is a direct measure of the impulsive change in $\chi(\mathbf{q}=\textup{M})$ created by ultrafast electronic excitation, which manifests as a stiffening of $\omega_{\textup{T}}$ according to Eqn.~\eqref{eqn:phonon_renorm}. This behavior arises from impulsive changes to the electronic distribution functions found in Eqn.~\eqref{eqn:struc_chi} following photo-excitation (i.e. photo-doping carriers into the electron pocket, Fig. 1a) and is thus a purely electronic effect. The subsequent $643\pm110$ rise in diffuse intensity at M$_{\parallel}$, however, is due to a combination of the two effects; i) lattice heating governed by $g_{\mathbf{q}}$ and ii) `re-softening' of the phonon due to carrier scattering out of the electron pocket (as we show below) yielding a redistribution of the electronic distribution functions. This behaviour is in contrast to the K and M$_{\perp}$ points in FIG.~\ref{fig:time_traces} where only a slower ($\sim$1~ps) diffuse intensity increase associated with lattice heating (phonon emission at that wavevector) are observed.  We observe no other impulsive suppression of $I_1(\mathbf{q})$ at other regions of the Brillouin zone, suggesting a pronounced anisotropy of $\chi_0(\mathbf{q})$. Our density functional theory calculations (Appendix~\ref{app:DFT}) reproduce the phonon renormalization behaviour for the $\omega_{\mathbf{T}}$ phonon. Simulations of the differential $I_1(\mathbf{q})$ intensity for different mode frequencies show excellent agreement with the measured intensity presented in FIG.~\ref{fig:3_panel_ueds}~b).

The magnitude of the scattering intensity from the M--point soft-mode forms a peak roughly 10 times smaller than the nearby $\Gamma=(120)$ and $\Gamma=(210)$ Bragg peaks (Inset of FIG.~\ref{fig:3_panel_ueds}~a). The equilibrium M$_{\parallel}$ intensity is shown in FIG.~\ref{fig:Mpeak_results}~a) and can be reliably fit to a Gaussian lineshape function with an offset. The amplitude of the lineshape is proportional to $1/\omega_{\textup{T}}^{2}$ (Eqn.~\ref{eqn:I1_reduced}) at the earliest delay times ($\sim$~200 fs), before significant lattice heating (see Appendix~\ref{app:diffuse_intensity}), and $n/\omega_{\textup{T}}$ at later times. The background intensity offset provides a measure of the diffuse background from the entire lattice system at that $\mathbf{q}$. Fitting a Gaussian plus offset to the transient intensity at all time-points allows for the separation of these two distinct physical processes. The results are presented in FIG.~\ref{fig:Mpeak_results}~c.  The M--point soft mode peak intensity at time-delays of 400 fs and 5 ps are displayed in a) and b) respectively along with best-fits. The dynamics of the peak amplitude shows two times-scales: IRF limited (130 fs) increase ($\omega_{\textup{T}}^{2}$ renormalization), followed by a $\sim700~$fs recovery of intensity. The diffuse background describing the lattice heating increases according to a single $\sim 1~$ps time-scale, consistent with the total intensity dynamics directly measured at K, M$_{\perp}$ and $\Gamma$ in FIG.~\ref{fig:time_traces}. Photo-carrier dynamics in 1T-TiSe$_2$ have been directly measured by time and angle-resolved photo-electron spectroscopy~\cite{Monney2016,Rohde2014} (tr-ARPES). These studies have determined the lifetime of the photo-doped carriers in the electron pocket at the M-point, which is shown in FIG.~\ref{fig:Mpeak_results}~c (black line) along with the results for fitting the M$_{\parallel}$ peak. These carrier dynamics cause $\chi(\mathbf{q}=\textup{M})$ to relax back towards its equilibrium value. Based on the trARPES results, it is expected that the quasi-impulsive, stiffening of $\omega_{\textup{T}}$ is followed by a `re-softening' on the observed carrier relaxation timescale.  This feature is reproduced in our extraction of the amplitude of the phonon peak at M$_{\parallel}$. The analysis presented in FIG.~\ref{fig:Mpeak_results} directly determines the frequency renormalization component of the change in soft-mode scattering amplitude. This renormalization leads to a peak amplitude change of ~40\%, significantly larger than that shown in FIG.~\ref{fig:time_traces} ($\sim1$\%) when examined in this manner because the intensity offset from the scattering due to other modes has been subtracted. For the incident fluence of 4 mJ/cm$^2$ applied in these experiments, we estimate a photo-carrier density of $\sim1\times10^{21}$cm$^{-3}$, corresponding to $\sim6\%$ excited unit-cells on average.

\begin{figure}[t!]
        \includegraphics[width=0.9\columnwidth]{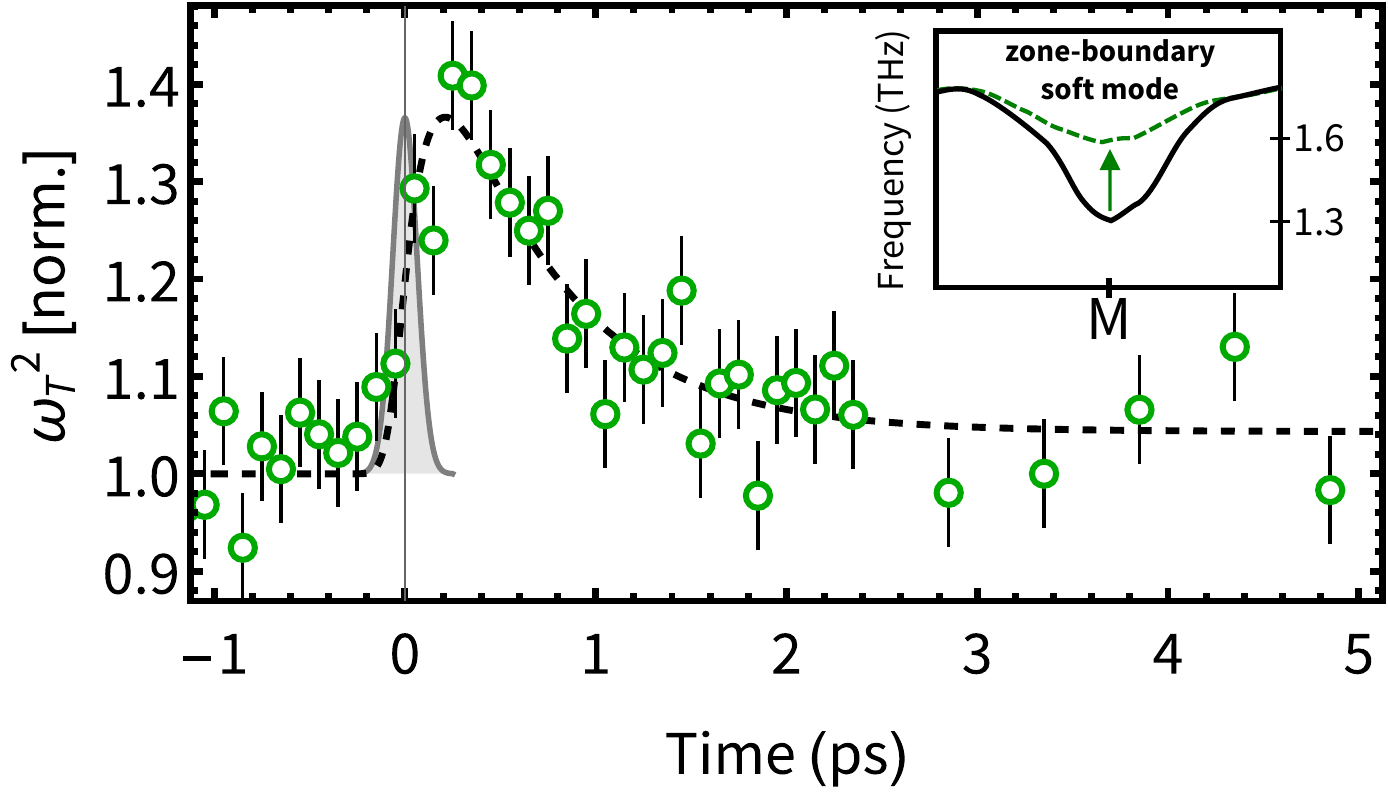}
        \caption{Magnitude of frequency renormalization for the soft transverse mode in TiSe$_2$ determined by transient scattering intensities at M--points (see Appendix~\ref{app:phonon_hardening_analysis} for details).  The gray Gaussian curve depicts instrumental temporal response function. The inset is a schematic representation of the change in phonon band for the soft-mode.}\label{fig:renormalization}
\end{figure}

A complimentary analysis can be performed to identify the phonon renormalization component directly from the UEDS intensity under the assumption that $g_{\mathbf{q}}$ is relatively constant throughout the Brillouin zone as our results strongly suggest (FIG~\ref{fig:3_panel_ueds} and Appendix~\ref{app:debye_waller_iso}). By subtracting the heating component from other phonons (\textit{not} the transverse mode which stiffens) determined from the UEDS data at M$_{\perp}$ where $\mathbf{q}=(0,\frac{5}{2},0)$ (Appendix~\ref{app:phonon_hardening_analysis}). At this $\mathbf{q}$, $F_{1j}(\mathbf{q})$ has essentially zero contribution from the transverse soft mode. With these lattice heating components removed, we obtain an estimate of the ``pure" phonon frequency re-normalization component of the UEDS signal measured at M$_{\perp}$.  We present $(\Delta\omega_{\mathrm{T}}/\omega_{\mathrm{T}})^2$ obtained this way in FIG.~\ref{fig:renormalization}. This complimentary analysis of the data shows a resolution-limited (130 fs) phonon stiffening followed by a $629\pm247~$fs softening, in agreement with the results of FIG.~\ref{fig:Mpeak_results}~c. This provides firm evidence that the observed phonon frequency renornalization is directly related to the carrier density in the electron pocket and such an effect is fundamentally different from the scattering found at all other regions of reciprocal space. 

\begin{table}[h]\caption{Electron-lattice equilibration rates.}\label{tab:rates}
\begin{center}
\begin{tabular}{c c c }
\centering
momentum $\mathbf{q}$ & time-constant (fs) & rate (THz)  \\
\hline
\hline
$\Gamma$ & $1070\pm30$ & $0.93\pm0.27$  \\
K        & $1270\pm190$ & $0.79\pm0.12$   \\
M$_{\perp}$ & $980\pm290$ & $1.03\pm0.31$    \\
M$_{\parallel}$ ($\omega_{\textup{T}}$--soft mode)& $1170\pm300$ & $0.88\pm0.21$  \\
\hline
\hline
\end{tabular}
\end{center}
\end{table}

\section{Discussion}\label{sec:discussion}

The data presented leads directly to the conclusion that the electron-phonon coupling vertex ($g_{\mathbf{q}}$) is not significantly enhanced for the soft zone-boundary phonons.  In 1T-TiSe$_2$, $g_{\mathbf{q}}$ is (to a good approximation) wavevector \textit{independent}, while the photoinduced modulation of $\chi(\mathbf{q})$ is strongly wavevector \textit{specific}. The time constants and rates are summarized in Table~\ref{tab:rates}. The ability to unravel these components of electron-phonon coupling is significant for the physics of CDW phases more generally, in particular the apparent competition with superconductivity~\cite{Calandra2011}, and is possible because of the combined momentum and time-resolution of UEDS. 

Our measurements reveal that photo-doping free carriers into the electron-pocket of TiSe$_2$ selectively `decouples' the $\omega_{\mathrm{T}}$ phonon mode, stiffening the vibration, \textit{i.e.}, $\omega_{\mathrm{T}}$ is directly correlated with free carrier density. However, in striking contrast to graphite~\cite{Stern2018} where the phonon modes that exhibit Kohn anomalies are also those into which electronic excitation energy flows most rapidly due to a strongly enhanced $g_\mathbf{q}$, in TiSe$_2$ there is no evidence of such an anisotropy in $g_\mathbf{q}$. From the perspective of the rate at which energy is transferred between free carriers and phonons, no strongly coupled phonon modes were observed. The soft phonon at the M-point does not exhibit an enhanced $g_\mathbf{q}$ and is not strongly coupled to free carriers in this sense. This provides complementary information to previous time- and angle resolved photo-electron spectroscopy measurements of semi-metallic TiSe$_2$~\cite{Monney2016}. These studies observed that photo-carrier doping at the $\Gamma$-point leads to an impulsive modification of the photo-emission signal at the M-point that was interpreted as being due to a disruption of excitonic correlations by excess free carriers. Our work shows that re-normalization of the transverse soft mode frequency also accompanies the injection of free carriers into the electron pocket, establishing a direct relationship between the energy of the soft mode and dielectric screening by free carriers.

Previous calculations of the bare susceptibility in TiSe$_2$ show weak divergences (enhancements) at $\Gamma$ and along M-L~\cite{Calandra2011,Kaneko2018}. There are two potential explanations for the observed weakening of the M-L divergence (\emph{i.e.} stiffening of the zone boundary phonons) resulting from photodoping free carriers into the electron pocket. First, photodoping yields a straightforward modulation of the carrier distribution functions (Eqn.~\eqref{eqn:struc_chi}), and this increased carrier density suppresses $\chi_0(\mathbf{q}=\textup{M})$. Second, the photodoped carriers enhance the dielectric screening of excitonic interactions~\cite{Kaneko2018,Monney2015} that are present in the system before photoexcitation.  Such correlations have been identified as important in TiSe$_2$, as they can govern the structure of the electronic bands near the Fermi-level and could be responsible for a substantial portion of the divergence in $\chi_0(\mathbf{q})$ at M even at room temperature.  We are not in a position to distinguish between these two possibilities in the current study.  Both of these explanations are qualitatively consistent with our measurements and the temperature dependence of the softening observed previously~\cite{Holt2001}. However, we propose that a subsequent temperature and photocarrier density (excitation fluence) dependent study could potentially distinguish between these possibilities when combined with theoretical predictions, in particular if complementary trARPES data were also available. It is also interesting to note that photodoping does not have a pronounced effect on the divergence in the bare susceptibility predicted to be present at $\Gamma$~\cite{Kaneko2018}, since no similar impulsive renormalization of phonon frequencies around $\Gamma$ is observed. This effect appears to be restricted to the soft zone boundary transverse phonons.

The observation that the electron-phonon vertex $g_{\mathbf{q}}$ in TiSe$_2$ is approximately constant as a function of wavevector is consistent with the dominant coupling between lattice displacements and the energy of electronic states being local in space ~\cite{Calandra2011}. That is, it is the phonon modulation of nearest neighbour Ti--Se distances that is critical to this coupling; the energy of the electronic states near the Fermi-level is sensitive primarily to the local chemical environments (orbital overlap) between nearest neighbour Ti and Se atoms.  For the case of the soft-phonon it has been previously proposed~\cite{Calandra2011} that the dispersion and softening is almost exclusively the result of the Ti-$3d$ chemical environment.

\section{conclusions}
In conclusion, we have applied UEDS to separate distinct contributions to momentum-dependent electron-phonon coupling in a complex material.  UEDS signals are profoundly sensitive to the photoinduced modulation of lattice structural fluctuations along phonon coordinates at all wave vectors, and naturally separate effects due to phonon frquency renormalization from those due to mode-dependent phonon heating. Thus, UEDS measurements are profoundly complementary to EELS measurements, which are sensitive to $\Im\left[\chi(\mathbf{q})\right]$, and ARPES measurements, which directly probe the occupancy of electronic states.  Our results demonstrate that the electron-phonon coupling vertex is relatively isotropic in the Brillouin zone of TiSe$_2$. This suggest that local interactions between nearest neighbour Ti and Se atoms are the dominant contribution to $g_{\mathbf{q}}$. By contrast, free carrier density in the electron pocket is found to govern the frequency of the soft-mode involved in the CDW transition in TiSe$_2$. A highly anisotropic electronic susceptibility strongly dependent on electron pocket free carrier density is the dominant mechanisms driving the temperature dependent phonon softening observed in measurements of TiSe$_2$ at equilibrium. 

\section*{acknowledgments}
This work was supported by the Natural Science and Engineering Research Council of Canada (NSERC), the Canada Foundation for Innovation (CFI) and Fonds de Recherche du Qu\'ebec--Nature et Technologies (FRQNT). J.-H. P. acknowledges the FRQNT PBEEE postdoctoral fellowship. This research was enabled in part by support provided by Calcul Quebec (https://www.calculquebec.ca/en/) and Compute Canada (www.computecanada.ca). The authors thank Jeannie Mui and the McGill Facility for Electron Microscopy Research (FEMR) for preparing the samples.

\section*{Author contributions}
B.J.S. and M.S. conceived the experiment. M.R.O. performed the experiments with help from L.P.R.de C., M.J.S. and J.-H.P. M.R.O. analyzed the data. J.-H.P. performed supporting DFT calculations. M.R.O. and B.J.S wrote the manuscript. M.R.O., J.-H.P. and L.P.R.de C. wrote the supplementary information.  All authors discussed the results and revised the manuscript.


\appendix

\setcounter{equation}{0}
\setcounter{figure}{0}
\setcounter{table}{0}
\renewcommand{\theequation}{A\arabic{equation}}
\renewcommand{\thefigure}{A\arabic{figure}}

\section{Experimental Methods \label{app:experimental_details}}

\subsection{Ultrafast electron scattering experiment}
In these experiments, the 90 keV bunch charge per pulse was roughly 5$\times 10^5~e^-$ and the temporal resolution was determined to be $\sim130~$fs~\cite{Otto2017} prior to the measurements. The repetition rate of the experiment was 1 kHz and scattered electrons were collected by a Gatan Ultrascan 1000 detector during a 10 second exposure. 48 idential pump-probe delay scans where carried out over the course of roughly 18 hours and averaged together with no time-zero correction applied in post-processing. The bulk 1T--TiSe$_2$ flakes were purchased from HQ Graphene and the sample was prepared by ultra-microtome at the McGill facility for electron microscopy research to a thickness of $70\pm10~$nm and placed over an TEM substrate (5 nm amorphous carbon) consisting of a 200 \textmu m radius aperture yielding an effective sample area of roughly 1.3$\times~10^{5}$ microns squared. The sample is excited with 1.55 eV (800 nm) and 50 fs (FWHM) laser pulses focused to 450~\textmu m (FWHM).

\subsection{Data processing and analysis}
Ultrafast electron scattering data is processed and analysed using the free and open-source program \texttt{iris}~\cite{RenedeCotret2018}, which is built on top of the \texttt{scikit-ued} and \texttt{npstreams} Python libraries.

\section{Properties of Titanium diselenide \label{app:tise2_properties}}

\begin{figure}[t]
        \includegraphics[width=0.25\textwidth]{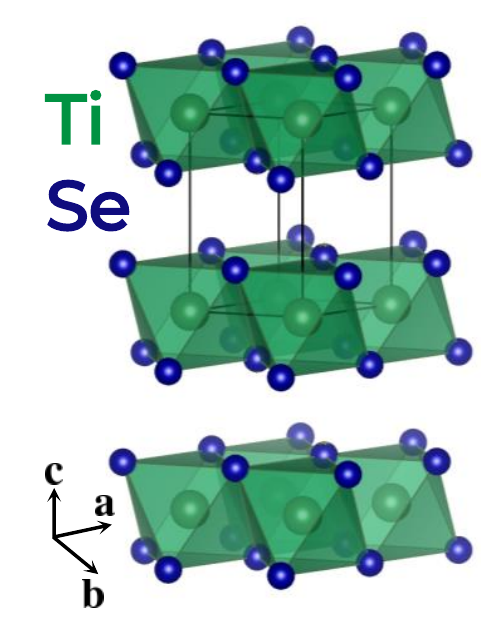}
        \caption{Crystal structure of the high-temperature phase of TiSe$_2$. Ti atoms are surrounded by Se octahedra, adjacent layers are bound by van der Waals forces.}\label{fig:TiSe2_crystal}
\end{figure}

1T--TiSe$_2$ crystallizes in the layered trigonal structure consisting of TiSe$_6$ octahedra (see Figure~\ref{fig:TiSe2_crystal}). 1T--TiSe$_2$ has a low-temperature phase (space-group: $P3c1$ [158]) and a high-temperature phase (space-group: $P\bar{3}m1$ [164]). The low-temperature phase is a 2$\times$2$\times$2 superstructure of the high-temperature phase. The high-temperature phase of TiSe$_2$ is semi-metallic~\cite{Zunger1978} and the electronic band structure at the Fermi energy is formed by partially empty Se--4p valence bands at $\Gamma$ and partially occupied Ti--3d conduction bands at M and L of the Brillouin zone.  In the semi-metallic phase, the octahedra are ordered, while they are distorted in the low-temperature phase resulting in a commensurate 2$\times$2$\times$2 charge-density wave system. The general effect of octahedral distortion leads to the back-folding of electronic band states (and therefore to the opening of a band gap).  The lattice constants are $a=b=3.54~$\AA~ and $c=6.01~$\AA~ determined by the manufacturer.  

\section{Isotropy of the transient Debye-Waller effect \label{app:debye_waller_iso}}

\begin{figure*}[t!]
        \includegraphics[width=1.75\columnwidth]{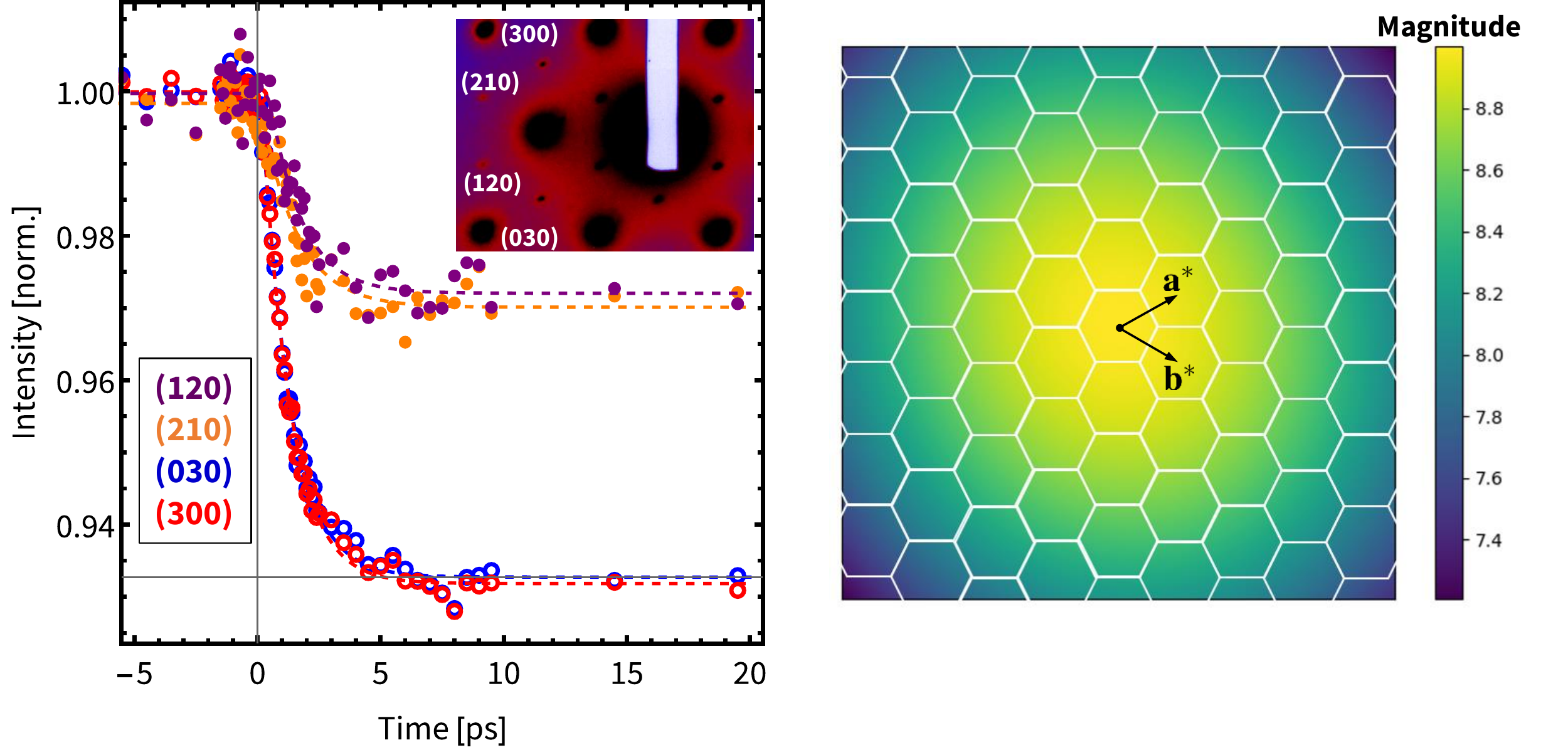}
        \caption{\textbf{Left:} Transient Debye-Waller effect along different reciprocal lattice directions measured by ultrafast electron diffraction.~\emph{Inset:} Portion of the TiSe$_2$ diffraction patten indicating the (300), (210), (120) and (030) Bragg peaks. \textbf{Right:} Computation of the anisotropic Debye-Waller factor $\sum_s W_s(\mathbf{q})$ using density functional theory (see Appendix~\ref{app:DFT}) for TiSe$_2$ showing no significant anisotropy.}\label{fig:DW_isotropy}
\end{figure*}

We observe in our measurements that the transient Debye-Waller effect is to a good approximation isotropic in momentum space. Figure~\ref{fig:DW_isotropy} shows four traces comparing $\mathbf{q}=(300)$ with $\mathbf{q}=(030)$ and $\mathbf{q}=(210)$ with $\mathbf{q}=(120)$. For $\mathbf{q}=(300)$ and $\mathbf{q}=(030)$ reflections we find time constants of $1.15\pm0.03~$ps and $1.17\pm0.04~$ps respectively.  For the $\mathbf{q}=(210)$ with $\mathbf{q}=(120)$ reflections we find time constants of $1.45\pm0.21~$ps and $1.32\pm0.22~$ps respectively.  As seen in FIG.~\ref{fig:DW_isotropy}, the magnitude of the intensity suppression is nearly identical at the same $|\mathbf{q}|$ in either the $\mathbf{a}^*$ or $\mathbf{b}^*$ directions. 

\section{Diffuse scattering intensity \label{app:diffuse_intensity}}

The first-order equilibrium thermal diffuse scattering (TDS) intensity at $\mathbf{q}$ is given by

\begin{eqnarray}\label{eqn:I1}
I_1(\mathbf{q}) &\propto& \sum_{j}\frac{|F_{1j}(\mathbf{q})|^2}{\omega_{j}(\mathbf{q})}\coth\left(\frac{\hbar\omega_{j}(\mathbf{q})}{2k_B T_j}\right)\\
&\approx & \sum_j\frac{n_j}{\omega_{j}(\mathbf{q})}|F_{1j}(\mathbf{q})|^2,
\end{eqnarray} 
where $\omega_{j}(\mathbf{q})$ is the phonon frequency and $T_{j}$ is the effective temperature of phonon mode $j$. More generally, the occupancy of phonons in branch $j$ is denoted by $n_{j}$. Equation~\eqref{eqn:I1} describes the contribution of all phonons to the scattering intensity. $F_{1j}(\mathbf{q})$ is the one-phonon structure factor which is given by~\cite{Xu2005}
\begin{equation}\label{eqn:1phonon}
|F_{1j}(\mathbf{q})|^2 = \left| \sum_s \exp\left(-W_s(\mathbf{q})\right)\frac{f_s(\mathbf{q})}{\sqrt{\mu_s}}\left(\mathbf{q}\cdot\hat{\mathbf{e}}_{j,s,\mathbf{k}}\right) \right|^2, 
\end{equation}
where $W_s(\mathbf{q})$ is the Debye-Waller factor, $f_s(\mathbf{q})$ is the atomic form for atom $s$ and $\hat{\mathbf{e}}_{j,s,\mathbf{k}}$ as the polarization vector of phonon $j$ with wavevector $\mathbf{k}=\mathbf{q-G}$; $\mathbf{G}$ is a reciprocal lattice vector and $\mathbf{q}$ is the scattering vector.

As indicated by Equation~\eqref{eqn:1phonon}, the one-phonon structure factors strongly depend on the product $\mathbf{q}\cdot\hat{\mathbf{e}}_{j,s,\mathbf{k}}$, the projection of the phonon polarization onto reciprocal space, $\mathbf{q}$.  The hyperbolic cotangent term in Equation Equation~\eqref{eqn:I1} is proportional to the phonon occupation. When $\hbar\omega_j(\mathbf{q})<k_BT_j$, Eqn.~\ref{eqn:I1} can be approximated as
\begin{equation}\label{eqn:I1_reduced_SI}
I_1(\mathbf{q})\approx\frac{2k_B}{\hbar}\sum_j\frac{T_j}{\omega_j(\mathbf{q})^2}|F_{1j}(\mathbf{q})|^2.
\end{equation}
This approximation is valid for phonons with frequencies less than $\sim6~$THz $\sim25$ meV at room temperature. In TiSe$_2$ this is always true for the soft phonon mode, which softens completely at the phase transition temperature, $T_c$, and stabilizes to roughly 3 THz at high temperature (see Appendix~\ref{app:DFT}).

\subsection{Intensity change at M$_{\parallel}$: Approximate fast dynamics}

Ultrafast electron diffuse scattering measures Eqn.~\eqref{eqn:I1} and~\eqref{eqn:I1_reduced_SI} in time. The normalized intensity change of the first-order thermal diffuse intensity is given by
\begin{eqnarray}
\Delta\bar{I}(\mathbf{q},t)&\equiv &\frac{I_1(\mathbf{q},t)-I_1(\mathbf{q},0)}{I_1(\mathbf{q},0)}  \\
&=& \frac{\sum_j\frac{n_j(t)}{\omega_{j}(\mathbf{q},t)}|F_{1j}(\mathbf{q})|^2-\sum_j\frac{n_j(0)}{\omega_{j}(\mathbf{q},0)}|F_{1j}(\mathbf{q})|^2}{\sum_j\frac{n_j(0)}{\omega_{j}(\mathbf{q},0)}|F_{1j}(\mathbf{q})|^2}\nonumber,
\end{eqnarray}
where we have assumed that $F_{1j}(\mathbf{q})$ does not change with time and that we have remained general in using $n_j(t)$ the phonons.  We expect only the transverse soft-mode in TiSe$_2$ (T) phonon to harden at $\mathbf{q}=~$M.  Inserting Eqn.~\eqref{eqn:I1_reduced_SI} for $j=$T and assuming that for the other modes ($j\neq~$T) $\omega_{j}(\mathbf{q},t)=\omega_j(\mathbf{q},0)$. This yields
\begin{eqnarray}\label{eqn:dIapprox}
&\Delta&\bar{I}(\mathbf{q}=\mathrm M,t) \approx \\ 
&\approx& \left(\frac{n_{\mathrm{T}}(t)}{n_{\mathrm{T}}(0)}\frac{\omega_{\mathrm{T}}(0)}{\omega_{\mathrm{T}}(t)} -1\right) + 
\frac{\sum_{j\neq\mathrm{T}}\frac{n_j(t)-n_j(0)}{\omega_{\mathbf{q}j}(0)}|F_{1j}(\mathbf{q})|^2}{\sum_{j\neq\mathrm{T}}\frac{n_j(0)}{\omega_{\mathbf{q}j}(0)}|F_{1j}(\mathbf{q})|^2}.\nonumber\\
&\approx& \left(\frac{T_{\mathrm{T}}(t)}{T_{\mathrm{T}}(0)}\frac{\omega_{\mathrm{T}}^2(0)}{\omega_{\mathrm{T}}^2(t)} -1\right) + \textup{~\sf~other phonon modes}.\nonumber
\end{eqnarray}

In obtaining Eqn.~\eqref{eqn:dIapprox}, we have assumed that $|F_{1\textup{T}}(\mathbf{q})|^2>|F_{1j}(\mathbf{q})|^2$ for all $j$, which is confirmed by our computational analysis (Appendix~\ref{app:DFT}). The first term in the above equation describes the intensity contribution from the T mode alone, in terms of both $T_{\mathrm{T}}(t)$ and $\omega_{\mathrm{T}}(t)$. The second term describes the change in phonon mode occupancies (proportional to $T_j$) undertaken by all phonons except T and physically contributes to the diffuse background.

The relationship between $\omega_{\mathrm{T}}$ and the electronic susceptibility $\chi(\mathbf{q})$ suggest a rapid timescale for photo-induced changes in $\omega_{\mathrm{T}}(t)$.  This is because $\chi_0(\mathbf{q})\propto f(\epsilon)$, the electronic distribution functions (Eqn.~\eqref{eqn:struc_chi}) which are effectively impulsively altered by photo-excitation~\cite{Kidd2002,Monney2011,Monney2016} as demonstrated by time-resolved ARPES experiments. This effect allows us to treat $T_{\mathrm{T}}(t)\approx~T_{\mathrm{T}}(0)$ as roughly constant over the time-scale during which $\omega_{\mathrm{T}}$ varies due to re-normalization (first 400 fs as evidenced by the data Ref.~\cite{Monney2016}). We assume that this is also true for the other modes given the times-scales present in the diffuse scattering data, suggesting that there a no phonon modes coupled strongly enough to compete with $\omega_{\mathrm{T}}(t)$. The phonon re-normalization component of $\Delta\bar{I}_{\mathrm{T}}(\mathbf{q}=\mathrm{M}_{\parallel},t)$ is then given simply by
\begin{equation}\label{eqn:hardening}
\Delta\bar{I}_{\mathrm{T}}(\mathbf{q}=\mathrm{M}_{\parallel},t) = -\left(1-\frac{\omega_{\mathrm{T}}^2(0)}{\omega_{\mathrm{T}}^2(t)}\right).
\end{equation}

The present work suggests a time-dependent $\chi_0(\mathbf{q}=\textup{M})$ following photoexcitation yields a re-normalization in $\omega_{\mathrm{T}}(t)$ relative to $\omega_{\mathrm{T}}(t\leq0)$ before photo-excitation. Making use of this we have
\begin{equation}\label{eqn:deltaIchi}
\Delta\bar{I}_{\mathrm{T}}(\mathbf{q}=\mathrm{M}_{\parallel},t) = -\left(1-\frac{\hbar\omega_0 - 2g^2_{\mathbf{q}}\chi_0(0)}{\hbar\omega_0-2g^2_{\mathbf{q}}\chi_0(t)}\right).
\end{equation}
Equation~\eqref{eqn:deltaIchi} describes the dynamics found in the results found in the main text.  Optical excitation reduces $\chi_0(t>0)$ producing a decreasing $\Delta\bar{I}_{\textup{T}}(\mathbf{q}=\textup{M}_{\parallel},t)$ because of transient hardening of the transverse phonon.  By treating photo-induced changes in $\chi_0(t)$ as small ($\chi_0(t)\approx \chi_0(0)-\Delta\chi_0(t)+\cdots$) we obtain
\begin{equation}\label{eqn:deltaI_chi_approx}
\Delta I(\mathbf{q}=\textup{M}_{\parallel},t) \approx - \frac{2g_{\mathbf{q}}^2\Delta\chi_0(t)}{\hbar\omega_0 - 2 g_{\mathbf{q}}^2\chi_0(0)}. 
\end{equation}

\section{Ultrafast electron scattering in the $\Gamma$--M--$\Gamma$--L plane \label{app:GMGL}}

\begin{figure}[t!]
        \includegraphics[width=0.5\textwidth]{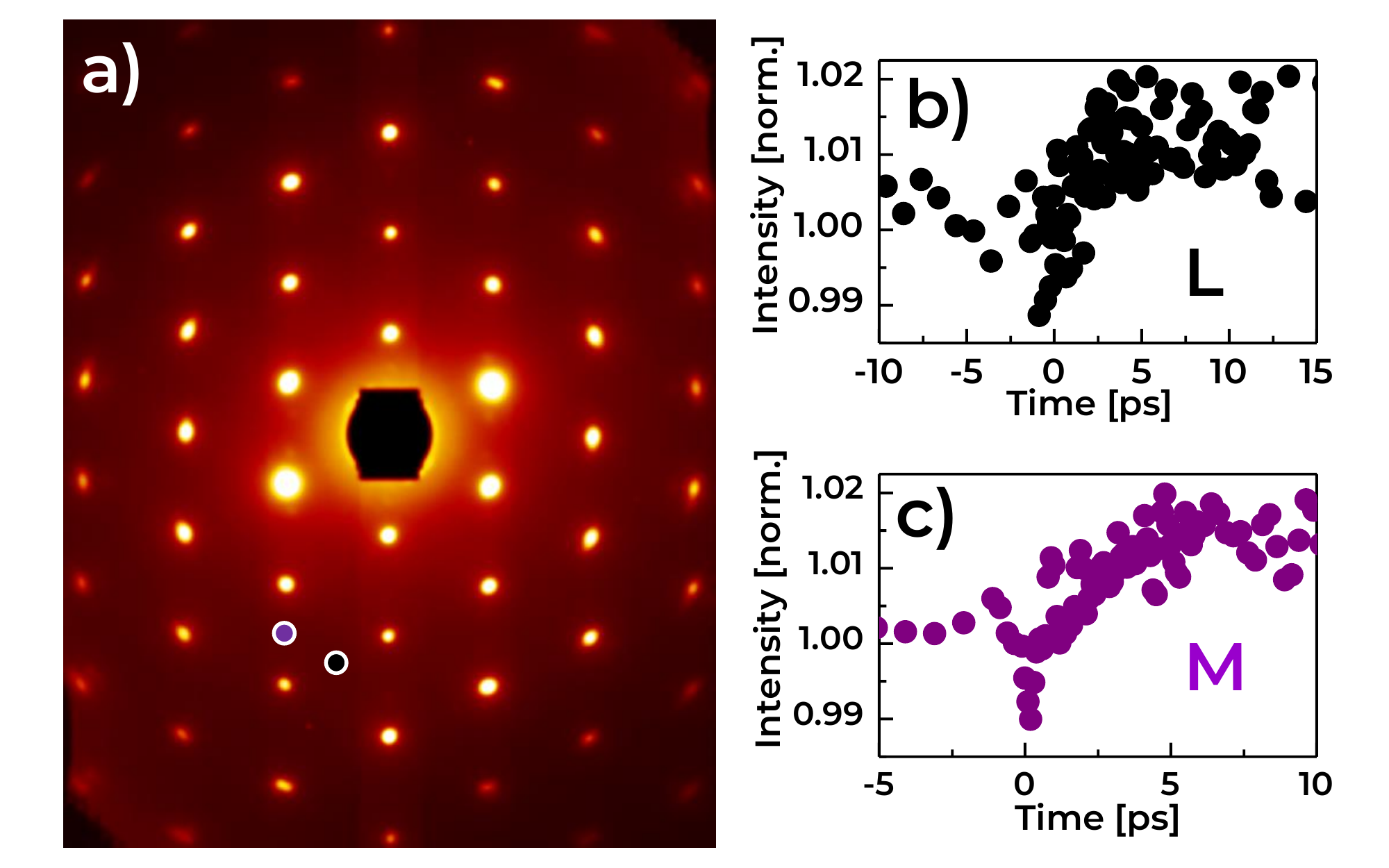}
        \caption{Ultrafast electron scattering in TiSe$_2$ at an angle corresponding to the $\Gamma$--M, $\Gamma$--L plane.  Time traces shown to the right exhibit the impulsive renormalization decrease described in the main text.}\label{fig:Lpoint}
\end{figure}

We also performed measurements with the TiSe$_2$ sample tilted at an angle to view the dynamics at the L point of reciprocal space.  We observe the same phonon softening behavior here, albeit with slightly reduced temporal resolution due to the velocity mismatch of the electron and optical pulses in this orientation. This is shown below in FIG.~\ref{fig:Lpoint}.

\section{Analysis of phonon renormalization \label{app:phonon_hardening_analysis}}
In this appendix we present complimentary analyses of the phonon renormalization effect described in section~\ref{sec:discussion}.

\subsection{Extraction from transient electron scattering intensity}

\begin{figure*}[t!]
        \includegraphics[width=0.85\textwidth]{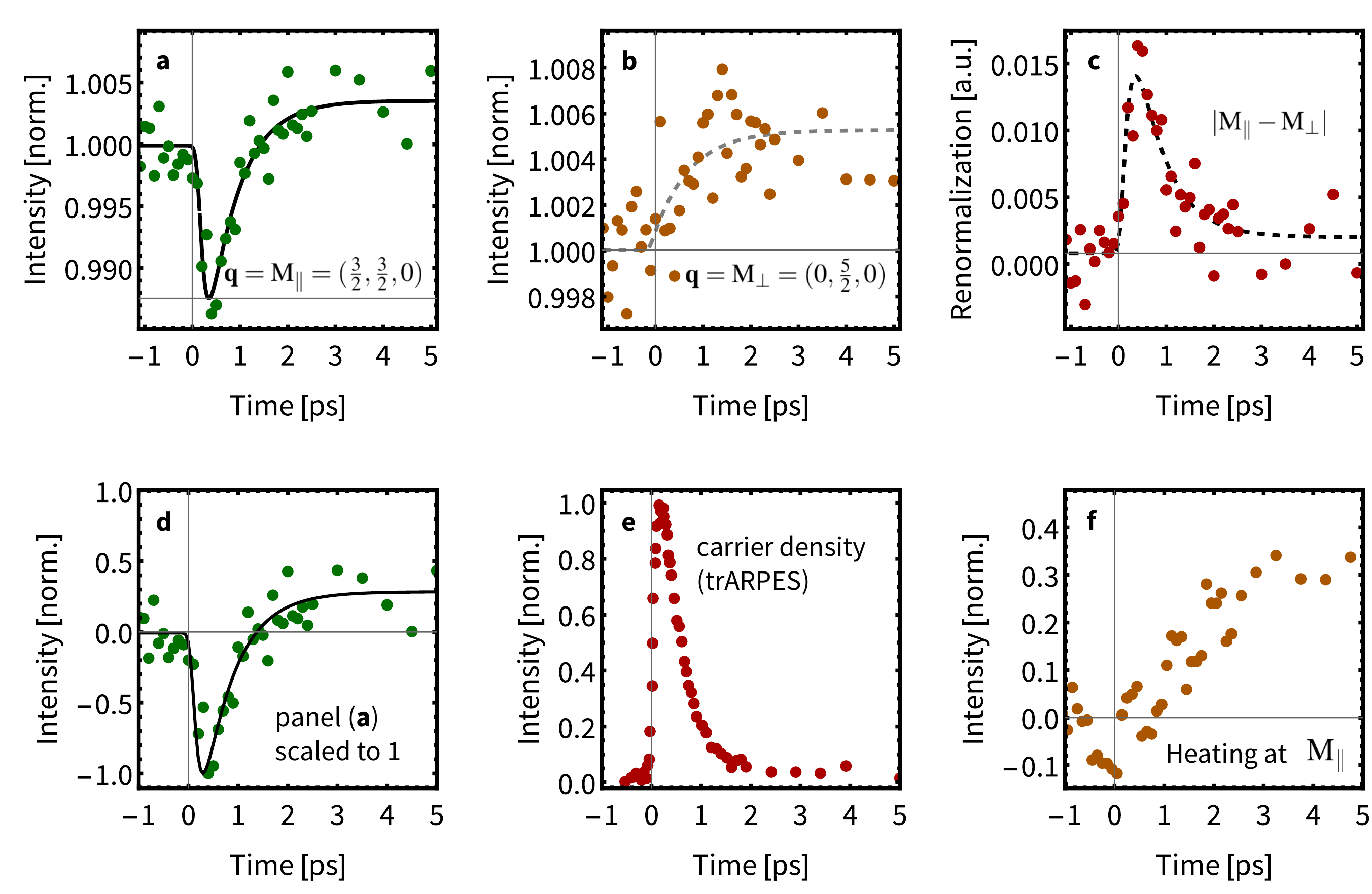}
        \caption{Extraction of phonon hardening component of ultrafast electron scattering intensity at $\mathbf{q=M}$. a) normalized intensity at $\mathbf{q}=\left(3/2,3/2,0\right)=\mathbf{M}_{\parallel}$ fit to a bi-exponential. b) Diffuse intensity rise at $\mathbf{q}=(0,5/2,0)$=M$_{\perp}$ where no rapid decrease is observed. c) Data from a) with data from b) subtracted (then multiplied by $-1$) along with a bi-exponential fit.  a) Ultrafast electron scattering data from the M$_{\parallel}$ point (also shown in Fig.~3a). b) Photo-excited electrons at M from Ref.~\cite{Monney2016}). c) Result of subtracting the traces from a) and b), corresponding to leftover phonon heating at M.} \label{fig:M_point_analysis}
\end{figure*}

The contribution of the phonon hardening induced change in diffuse scattering at M$_{\parallel}$ ($\mathbf{q}=\left(\frac{3}{2},\frac{3}{2},0\right)$, see FIG.~\ref{fig:M_point_analysis} a)) is isolated from the total signal by considering the diffuse rise found at another M--point in reciprocal space at a very similar $|\mathbf{q}|$. This M--point, M$_{\perp}$ at $\mathbf{q}=\left(0,\frac{5}{2},0\right)$, is shown in FIG.~\ref{fig:M_point_analysis}~b). This data represents an average time-scale for the increase in phonon occupancy at such a point due to lattice heating from electron-phonon coupling which we show earlier (Appendix~\ref{app:debye_waller_iso}) is isotropic to a very good approximation. This data is fit to a single exponential given by (fitting parameters given in Table ~\ref{tab:fitting})
\[
f(t-t_0) = \Theta(t-t_0)\times A \left(1 - \exp\left(-(t-t_0)/\tau\right)\right).
\]
This fit is then subtracted from the data shown in FIG.~\ref{fig:M_point_analysis}~a) yielding the data shown in c) (absolute value is shown to depict specifically the increase in phonon frequency).  This data is then fit to a bi-exponential function of the form
\begin{eqnarray*}
g(t-t_0) &=& \Theta(t-t_0)\times A\left(1-\exp\left(-(t-t_0)/\tau_1\right) \right) + \cdots \\
&+& \Theta(t-t_0)\times B\left(1-\exp\left(-(t-t_0)/\tau_2\right)\right) ,
\end{eqnarray*}
for which the fitting results are shown as the dashed line. This response describes both the fast initial impulsive hardening followed by the slower softening of the phonon with the heating contribution approximately removed. For a point of comparison, we may also fit the total UEDS intensity dynamics at M$_{\parallel}$ to the same mode and compare the results.  The best-fit parameters are shown in Table~\ref{tab:fitting}.

\begin{table}[ht]
\caption{Best-fit parameters for the diffuse intensity dynamics at M$_{\parallel}$ and M$_{\perp}$.}
\begin{tabular}{|c|c|c|c|}
\hline
parameter & \shortstack{total M$_{\parallel}$ data \\ FIG.~\ref{fig:M_point_analysis}~a} & \shortstack{phonon component  \\  FIG.~\ref{fig:M_point_analysis}~c} & \shortstack{total M$_{\perp}$ data \\ FIG.~\ref{fig:M_point_analysis} b}  \\
\hline
\hline
$A$       & $-0.024\pm0.004$ & $0.027\pm0.017$ & $5.2\pm0.5\times10^{-3}$               \\
$B$       & $0.027\pm0.004$ & $-0.026\pm0.017$ & --              \\
$\tau_1$  & $107\pm21$ fs & $152\pm122$ fs & --               \\ 
$\tau_2$  & $643\pm110$ fs & $629\pm247$ fs & $757\pm334$ fs             \\
$t_0$     & $56\pm21$ fs & $24 \pm28$ fs & $-64\pm105$ fs             \\ \hline
\end{tabular}
\label{tab:fitting}
\end{table}

\subsection{Extraction of remaining phonon heating using time/angle-resolved photo-electron spectroscopy data}

\begin{figure*}[t!]
        \includegraphics[width=1.9\columnwidth]{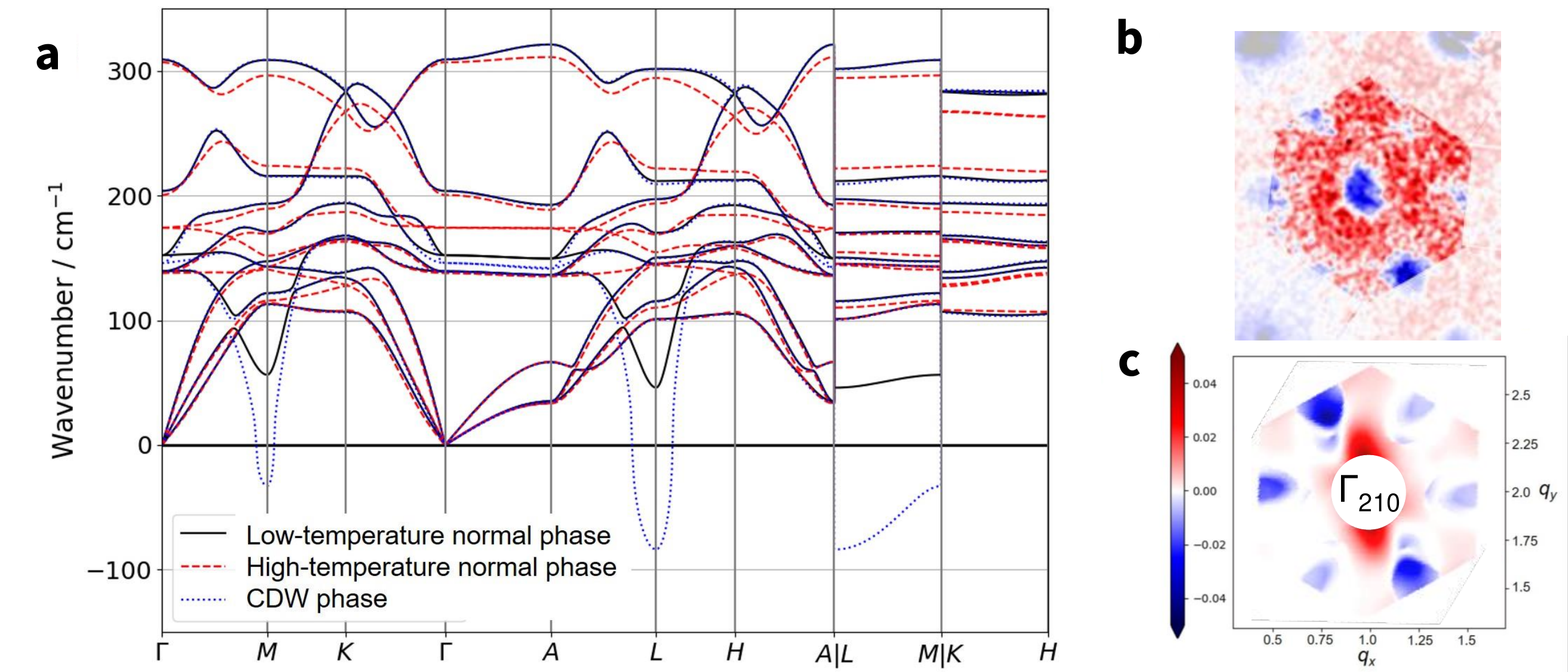}
        \caption{Comparison of experimental diffraction data with density functional theory calculations. (a) Computed phonon dispersion curve of TiSe$_2$ in the charge-density wave (CDW) phase and normal phase at two different temperatures showing the renormalization behavior of the zone-boundary transverse mode (dotted blue line at M and L) and the flat dispersion character between M--L symmetry points. (b) Experimental (from FIG~\ref{fig:3_panel_ueds}~b) and (c) computation of the intensity difference in diffuse scattering in the Brillouin zone around the (210) peak. }\label{fig:DFT}
\end{figure*}

The data from the M$_{\parallel}$ point are shown in FIG.~\ref{fig:M_point_analysis}~a) and d) and consists of two distinct components, a fast phonon-hardening component, and a slower diffuse intensity rise from all phonons.  Monney \textit{et al.} \cite{Monney2016} measured the time-dependence of the photo-excited carrier distribution at the M--point of the electronic band structure which is shown in FIG.~\ref{fig:M_point_analysis}~e).  Although they excite with 1300 nm and hence, drive vertical transitions at $\Gamma$ rather than directly at M, their results show that the excited electrons quickly (~10--50 fs) populate the nearby band minima located at M.  Our results indicate that these photo-excited carriers play a role in hardening the phonon frequency through the suppression of charge screening effects (and equivalently the reduction of the susceptibility $\chi$. We can remove the normalized photo-excited electron intensity from the ultrafast electron diffuse scattering data from the M$_{\parallel}$ (FIG.~\ref{fig:M_point_analysis}~d)) region of the pattern.  What remains of the signal is a $\sim1~$ps time-scale (see FIG.~\ref{fig:M_point_analysis}~f)) rise which agrees well with the ultrafast electron scattering data at the M$_{\perp}$ points where no rapid decreasing intensity (renormalization) is found (this also agrees with average transient Debye-Waller effect observed on the Bragg peaks). We can only glean a qualitative comparison of the timescales here since the magnitudes of these signals will manifest differently because of differences between the electron scattering and photo-electron spectroscopy. 

\subsection{Electron-phonon coupling upper bound}

We justify our claim that electron-phonon coupling must be relatively isotropic compared to $\chi_0(\mathbf{q})$ in the following manner: The $\sim650$ fs rise at M$_{\parallel}$ (FIG.3~a) from main text) is $\sim30\%$ faster than the $\sim1$ ps decrease found at the $\Gamma$--points and $\sim 65\%$ faster than the $\sim1$ ps rise found at K and M$_{\perp}$.  This establishes that electron-phonon coupling is \textit{at most} 65\% stronger at M. This difference (or small anisotropy) is much less than the order of magnitude difference between the 100 fs phonon renormalization due to $\chi_0(\mathbf{q})$ (via the modulation of electronic states) and the electron-phonon coupling rate at the same wave-vector.  This upper-bound does not account for the fact that the phonon must re-soften (Appendix~\ref{app:phonon_hardening_analysis}).  Unaccounted for, this re-softening adds an additional fast component to the rise at M$_{\parallel}$.

\section{Computational Methods \label{app:DFT}}

The computed phonon dispersion curve of TiSe$_2$ indicates that the transverse soft mode phonon exhibits an imaginary frequency at the M and L points in the CDW phase (Fig.~\ref{fig:DFT} (a)), which has been reported elsewhere~\cite{Fu2016,Duong2015}. This is consistent with the additional Bragg peaks at the M and L points in the diffraction pattern of the CDW phase. In the normal phase, the transverse soft mode hardens with increasing temperature above the CDW phase transition until its maximum frequency is reached (see FIG.~\ref{fig:DFT}~(a)). In addition to the TA phonon mode, there is a slight change of the TO mode at the high-symmetry line $\Gamma$--M. 

To investigate the effects of phonon hardening with temperature in relation to the intensity change of the diffraction pattern, we calculated the change in intensity given by $I_1\propto \sum_{j,\mathbf{q}}\frac{1}{\omega_{j,\mathbf{q}}}\left(\sum_s\mathbf{q\cdot\hat{e}}_{j,\mathbf{q},s}\right)^2$ at the $\Gamma_{120}$ zone from the phonon dispersion curves obtained for room temperature and high-temperature. The difference between both intensities (high-temperature subtracted from room temperature) is compared to the experimental differential diffraction data (FIG.~\ref{fig:DFT}~(b) and (c)). While the atomic factor does not contribute to the relative change in the intensity, the Debye-Waller factor increases at the $\Gamma$--point leading to a reduction in intensity and an increase in intensity in the remaining Brillouin zone.  The computed diffuse diffraction pattern agrees well with the experiment. Figure~\ref{fig:DFT}~(c) shows that both computed M$_{\parallel}$ points at $\left(3/2,3/2,0\right)$ and $\left(1/2,5/2,0\right)$ reveal a stronger intensity reduction than at the M$_{\perp}$ points. Furthermore, the M point at (1/2,2,0) is predicted to be more intense than the corresponding M point at (3/2,2,0) which is consistent with the experimental pattern (Fig.~\ref{fig:DFT}~(b)). 

\subsection{Density functional theory approach}
The phonon dispersion curves including the phonon frequencies and polarization vectors of TiSe$_2$ were computed using the PHonon package in Quantum ESPRESSO~\cite{Giannozzi2009} with the B86b exchange-coupled~\cite{Becke1986} Perdew-Burke-Ernzerhof (B86bPBE) generalized gradient approximation (GGA)~\cite{Perdew1996} and the projector augmented-wave (PAW) method. The cut-off energy of the wavefunction (charge density) was converged prior the relaxation leading a cut-off energy of the wavefunction (charge density) of 82 Ry (972 Ry). 

TiSe$_2$ is a layered structure and to include the dispersion forces between the layers, the exchange-hole dipole moment (XDM) method was implemented~\cite{Becke2007}. Prior to the phonon calculations, the crystal structure was fully relaxed in two steps. First, the crystal structure was relaxed using a $\Gamma$--centered 24$\times$24$\times$12 $k$-point mesh, force (energy) threshold of 10$^{-4}$ Ry Bohr$^{-1}$ (10$^{-1}$ Ry) and a Fermi-Dirac smearing of 10$^{-3}$ Ry. In the second step, the atomic positions were relaxed to a $\Gamma$--centered 16$\times$16$\times$16 $k$--point mesh, force (energy) threshold of 10$^{-7}$ Ry Bohr$^{-1}$ (10$^{-10}$ Ry) and a Fermi-Dirac smearing of 1.9$\times 10^{-3}$ Ry for the charge-density wave phase and 1.9$\times 10^{-2}$ Ry for the normal phase. The atomic positions were relaxed until the total force was equal to zero. The dynamic matrices were computed on 4$\times$4$\times$2 $q$--point grid using a self-consistency threshold of 10$^{-16}$ Ry. The phonon frequencies/polarization vectors in the triangle $\Gamma$-M-K were interpolated on a 100$\times$67 $k$-point mesh. 

To compute the change in intensity, the difference in $I_1\propto \sum_{j,\mathbf{q}}\frac{1}{\omega_{j,\mathbf{q}}}\left(\sum_s\mathbf{q\cdot\hat{e}}_{j,\mathbf{q},s}\right)^2$ of the normal phase at room temperature and high temperature was calculated. For the room-temperature phase, the minimum frequency was set to 53 cm$^{-1}$ (\textit{i.e.}, minimum frequency of the M-point at room temperature), in agreement with experimental~\cite{Holt2001} and computational data~\cite{Fu2016}.

\section{On the validity of the kinematical approximation \label{app:kinematic_simulations}}

\begin{figure}
    \centering
    \includegraphics[width=0.9\columnwidth]{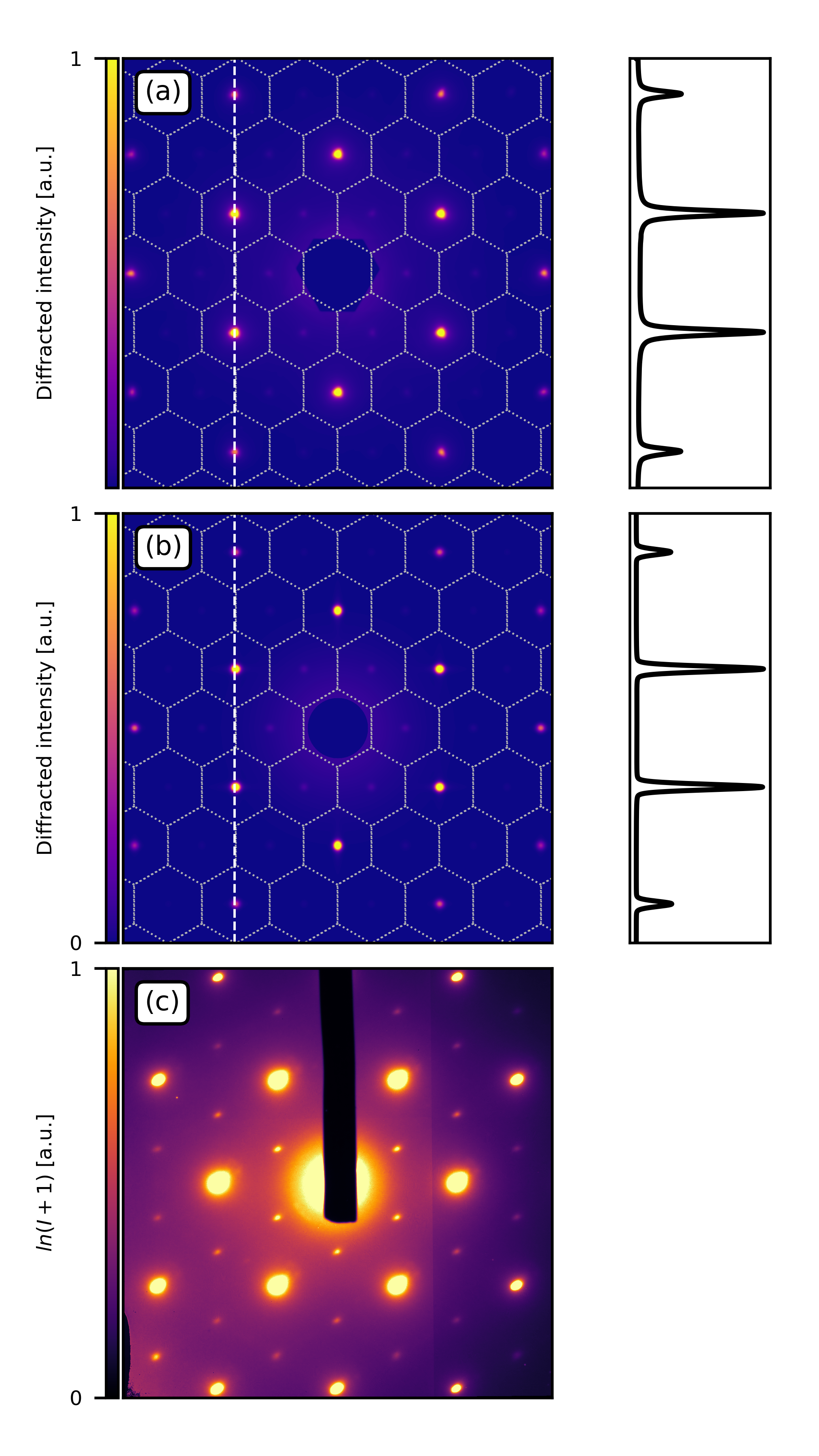}
    \caption{Comparison between (a) symmetrized, pre-photoexcitation electron diffraction pattern of TiSe$_2$ and (b) kinematical diffraction simulation of an 8x8 supercell. The Brillouin zones are marked by dashed hexagons. Equivalent line cuts from (a) and (b) are shown on the right of each subfigure. The location of the line cut is highlighted with a horizontal dashed line. This comparison indicates that the diffraction presented in this work is mostly kinematical. Panel (c) shows an unprocessed, pre-photoexcitation electron diffraction pattern where the contrast has been increased by taking the natural logarithm of the intensity ($I \mapsto \ln(I + 1)$). Even with this increased contrast, Kikuchi lines are not visible at all, indicating that multiple diffuse scattering effects are not significant.}
    \label{fig:kinematic_diff}
\end{figure}

Diffraction pattern simulations were performed in order to confirm that multiple scattering effects were not strongly present in the data. These simulations were performed following the weak-phase approximation developed in \citet{Kirkland2008} and implemented in \texttt{scikit-ued}~\cite{RenedeCotret2018}. The resulting simulated pattern for an 8x8 supercell is compared to pre-photoexcitation data visible on FIG.~\ref{fig:kinematic_diff}. The lack of significant multiple scattering is also evident from the absence of Kikuchi lines in the scattering patterns~\cite{Fultz2001}. FIG.~\ref{fig:kinematic_diff} c) shows a diffraction pattern of TiSe$_2$ before photoexcitation, where the contrast has been increased dramatically by taking the natural logarithm ($I \mapsto \ln(I + 1)$. Even with this increased contrast, Kikuchi lines are not visible at all.


\bibliography{bibliography}

\end{document}